\documentclass[twocolumn,aps,amsmath,amssymb,prb,showpacs]{revtex4}
\usepackage{epsfig,bm,dcolumn}
\usepackage{graphicx}

\begin{document}

\author{C. Menotti$^{1,2}$}
\author{N. Trivedi$^{3}$}
\affiliation{$^1$ ICFO - Institut de Ciencies Fotoniques,
Mediterranean Technology Park,
E-08860 Castelldefels (Barcelona), Spain \\
$^2$ CNR-INFM-BEC and Dipartimento di Fisica,
Universit\`a di Trento, I-38050 Povo (Trento), Italy\\
$^3$ Department of Physics, The Ohio State University,
Columbus, Ohio 43210, USA}

\title{Spectral weight redistribution in strongly correlated bosons in
optical lattices}

\begin{abstract}
We calculate the single-particle spectral function for the
one-band Bose-Hubbard model within the random phase approximation
(RPA). In the strongly correlated superfluid, in addition to the
gapless phonon excitations, we find extra gapped modes which
become particularly relevant near the superfluid-Mott quantum
phase transition (QPT). The strength in one of the gapped modes, a
precursor of the Mott phase, grows as the QPT is approached and
evolves into a hole (particle) excitation in the Mott insulator
depending on whether the chemical potential $\mu$ is above (below)
the tip of the lobe. The sound velocity $c$ of the Goldstone modes
remains finite when the transition is approached at a constant
density, otherwise, it vanishes at the transition.  It agrees well
with Bogoliubov theory except close to the transition. We also
calculate the spatial correlations for bosons in an inhomogeneous
trapping potential creating alternating shells of Mott insulator
and superfluid. Finally, we discuss the capability of the RPA
approximation to correctly account for quantum fluctuations in the
vicinity of the QPT.
\end{abstract}


\maketitle

\section{Introduction}

Optical lattices have made it possible to explore the properties of
ultracold dilute atoms in a new regime of strong
correlations~\cite{optical_lattices,jaksch_prl1998,bloch_nature2002}. By
tuning the strength of the laser field, the effective interactions
between atoms can be tuned to become stronger than their kinetic
energy. For Bose systems, such a competition between kinetic $t$ and
interaction energy $U$ drives a quantum phase
transition~\cite{fisher_prb1989,sachdev_qpt} from a kinetic energy
dominated superfluid (SF) phase to an interaction dominated Mott
insulating (MI) phase. The Bose Hubbard model (BHM) captures the
essential physics of this problem~\cite{fisher_prb1989}, provided the
interactions between the bosons are smaller than interband energies
and the problem can be treated in the single band approximation. While
the BHM was proposed much before optical lattice experiments became
available, a direct experimental realization was missing. In condensed
matter systems, Josephson junction arrays~\cite{jjarrays}, $^4$He in
vycor and aerogels~\cite{he4}, vortices in
superconductors~\cite{vortices} and quantum magnets~\cite{auerbach}
can be modeled by the BHM, but the actual systems have additional
complications of disorder or longer range interactions which make the
comparisons between theory and experiment difficult.

One of the main advantages of the cold atom systems is that they
are clean and much more tunable: the density of bosons, their
effective interaction, the tunneling amplitude between the wells,
the number of lattice sites, the shape of the trapping potential
and aspect ratios can all be varied rather easily, making it
possible to study the effects of inhomogeneity and dimensionality.
In addition, it is possible to add random potentials to study the
effects of disorder. This sets the optical lattice systems apart
as a useful testing ground for theoretical ideas in the area of
strongly interacting bosons and fermions.  This model has also
provided tremendous impetus for the development of new measurement
techniques to address questions about the nature of the
excitations especially near the transition ~\cite{altman_prl02,
stoferle_prl2004, kramer_pra2005, vanoosten_pra2005,
sengupta_pra2005, nikuni, ohashi_pra2006, huber_prb2007, grimm}.
The recent experiments on the
dynamics~\cite{bloch2_nature2002,zakrzewski} have given a window
into the different time scales operating within the different
phases and around the quantum phase transitions.

The paper is organised as follows: In Sect.~\ref{sect_model}, we
present the Bose-Hubbard model and state our main results. In
Sect.~\ref{sect_rpa}, we discuss the nature of the spectral
function calculated within the RPA formalism as it evolves from
the SF to the MI phase by decreasing $t/U$. The sound velocity
in the SF phase and its comparison with Bogoliubov theory is
contained in Sect.~\ref{sect_c}. The momentum distribution and the
spatial correlations are calculated in Sect.~\ref{nqrhorr}. The
RPA formalism is generalized to a spatially inhomogeneous trapping
potential in Sect.~\ref{sect_inhom}. We conclude in
Sect.~\ref{conclusions} with some remarks about the comparison
between RPA and mean field theory. There are three appendices that
give the details of the calculations of the Green function within
RPA (App.~\ref{app1}), the Bogoliubov calculation for the
Bose-Hubbard model (App.~\ref{app2}), and the momentum
distribution function within RPA in the Mott regime
(App.~\ref{app3}).

\section{Model and Main Results}
\label{sect_model}

The Bose-Hubbard Hamiltonian is defined as

\begin{equation}
H=-\frac{t}{2z} \sum_{\langle i j\rangle} \left( a^{\dag}_i a_j +
a_i a^{\dag}_j \right) + \frac{U}{2}\sum_i n_i(n_i-1) - \mu \sum_i
n_i\ , \label{bhm}
\end{equation}
where $a_i$ and $a^{\dag}_i$ are bosonic annihilation and creation
operators respectively and $n_i=a^{\dag}_i a_i$ is the density
operator. The parameter $U$ describes the on-site repulsive
interaction between bosons, $t$ is the tunneling parameter between
nearest neighbors as indicated by the symbol $\langle i j\rangle$,
$\mu$ is the chemical potential that fixes the number of particles and
$z=2D$ is the coordination number in $D$ dimensions. This Hamiltonian
shows a quantum phase transition (QPT) from the SF to the MI phase as a
function of $t/U$. The theoretical approaches used to investigate this
Hamiltonian include mean field theory~\cite{sheshadri_epl1997},
perturbation theory~\cite{monien}, variational
methods~\cite{variational} and quantum Monte Carlo simulations
~\cite{scalettar_prl1990,krauth-nt}.

In this paper, we use a Green's function formalism to investigate
the excitations and correlations in the BHM. We start from the
mean-field (MF) ground state which is essentially a product of
single site states and go beyond it by including inter-well
coupling within a random phase approximation (RPA) ~\cite{haley,
zubarev}.

Our main results are as follows:

(1) In the weakly interacting SF ($t/U\approx 100$) the gapless
phonons of the SF, or the Goldstone modes arising due to the broken
gauge symmetry, exhaust the sum rule on the total spectral weight, as
expected. For $t/U\approx 10$ there are already small deviations from
Bogoliubov theory and new gapped modes appear in the SF phase. These
gapped modes pick up strength as $t/U\approx 1$. The sum rule is now
satisfied only upon including {\em both} phonon and gapped modes.

(2) At the transition, we observe the progression of one of the phonon
modes in the SF into a gapped mode in the MI (the one which is gapless
at the QPT). The second gapped mode in the MI instead arises from one
of the gapped modes in the SF. Such gapped modes in the SF have been
reported previously using several theoretical methods
~\cite{sengupta_pra2005,nikuni,ohashi_pra2006,huber_prb2007}. We argue
that these additional gapped modes are a distinctive signature of a
strongly correlated SF in proximity to a MI in an optical
lattice. They indicate the redistribution of spectral weight from the
coherent phonon modes into incoherent excitations, and are a precursor
of the MI beyond the QPT.

(3) We calculate the sound velocity in the RPA formalism and show that
it agrees with $c=1/\sqrt{\kappa m^\ast}$ calculated independently
from the mean-field effective mass and compressibility. In a wide
range of parameters, except very close to the SF-MI phase transition,
the above sound velocity compares well with the predictions of
Bogoliubov theory.

(4) We exploit a special feature of superfluids that allows us to extract the
condensate fraction $n_0$ from the strength of the phonon modes in
the spectral function.

(5) We calculate the spatial correlations in the case where an
inhomogeneous confining potential is superimposed on the optical
lattice. The response to a perturbation is strongly influenced by
the presence of alternating shells of Mott insulator and
superfluid regions.

\section{Formalism: RPA approximation, spectral function, and excitations}
\label{sect_rpa}

We start with the mean-field approximation in real
space~\cite{sheshadri_epl1997} obtained by giving the annihilation and
creation operators an expectation value defined by $\langle a\rangle
=\langle a^{\dag }\rangle =\varphi$. The order parameter $\varphi $
identifies the nature of the system: it is non-zero in the SF phase
and vanishes in the insulating phase. Substituting $a={\varphi+
\tilde{a}},\;\;a^{\dag }=\varphi+{\tilde{a}}^{\dag }$, in
Eq.~(\ref{bhm}), where $\tilde{a}$ and $\tilde {a}^\dag$ are the
fluctuations of the Bose field around the mean field value, the
Hamiltonian $H$ can be rewritten as a sum of on-site Hamiltonians

\begin{equation}
H_{i}^{MF}=\frac{U}{2}n_{i}(n_{i}-1)-\mu n_{i}-t\varphi
(a_{i}^{\dag }+a_{i})+t\varphi ^{2}, \label{bhm2}
\end{equation}
which include the tunneling at the mean field level through the order
parameter $\varphi$. In the MF approximation, we neglect the non-local
inter-well hopping term $-(t/2z) \sum_{\langle i j \rangle} \left(
{\tilde a}^{\dag}_i {\tilde a}_j + {\tilde a}_i {\tilde a}^{\dag}_j
\right)$, which we will later treat in RPA.

The $H_{i}^{MF}$ can be diagonalized numerically, leading to a set of
on-site eigenstates such that $H_{i}|i\alpha \rangle =\epsilon
_{\alpha }|i\alpha \rangle $. In the Mott limit the eigenstates
$|i\alpha \rangle $ are number states, while in the SF regime they are
coherent superpositions of several number states, allowing the order
parameter to be different from zero. The MF ground state solution is
given by the product state $|\Phi \rangle =\Pi_{i}|i,0\rangle $,
equivalent to the one obtained in the Gutzwiller Ansatz, where
$|i,0\rangle$ is the ground state of $H_i^{MF}$.

\begin{center}
\begin{figure}[h!]
\includegraphics[width=0.7\linewidth]{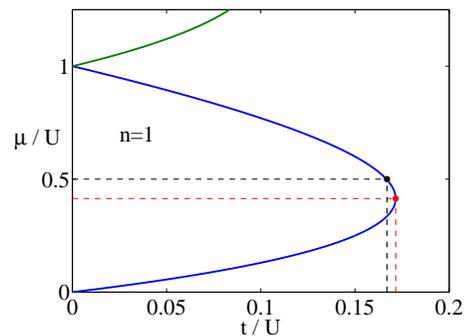}
\caption{(Color online) Mean field phase diagram in the $\mu/U$
{\it vs.} $t/U$ plane. The blue line shows the Mott insulating
lobe at density $n=1$. On this diagram, we indicate two points
where the QPT happens, that we are going to discuss in this paper:
a generic one at $\mu/U=0.5$ and $t/U \approx 0.167$ (black) and
the tip of the lobe at $\mu/U=\sqrt{2}-1$ and $t/U \approx 0.1716$
(red) where the QPT takes place at constant density.}
\label{phase_diag}
\end{figure}
\end{center}

Within the mean field approximation, the state of the system is
described by a product state over the different wells, neglecting all
inter-well correlations. However, even in the MI, correlations between
neighboring wells do not vanish; in fact they get large as $t/U$ is
increased from the Mott side, ultimately diverging at the
transition. Experimental evidence of these correlations is found in
the interference picture of an atomic cloud released from a 3D optical
lattice, the visibility of which does not suddenly vanish at the phase
transition~\cite{bruder, bloch_prl2005, gangardt, spielman_prl2007,
  diener_prl2007}. These important features are captured by the RPA
performed on the non-local tunneling terms of the BHM. At the end of
this paper, we discuss the limitations of the RPA method and how it
compares with the mean-field approximation.

To go beyond the mean-field approximation, we treat the inter-well
hopping term within RPA, as described in App.~\ref{app1}
\cite{haley,zubarev}. This method allows us to compute the Green's
function $ G({\bf q},\omega) = \langle\langle a^{\dag}_{\bf q} ;
a_{\bf q} \rangle\rangle_\omega$, defined in
Eqs.~(\ref{gtau},\ref{gomega}), and from that, the spectral
function ${\cal A}({\bf q},\omega)=-(1/\pi){\rm Im} G({\bf q},\omega)$.

Due to the commutation relations of the bosonic destruction and
creation operators, the spectral function always satisfies the sum
rule

\begin{eqnarray}
\int_{-\infty}^\infty \mathcal{A}({\bf q},\omega) d \omega = 1.
\label{sumrule}
\end{eqnarray}
From the spectral function, one can extract the excitation
spectrum, the strength of the excitation modes, and the related
density of states

\begin{eqnarray}
{\rm DOS}(\omega)=\int {\cal A}({\bf q},\omega) d{\bf q}.
\label{dos}
\end{eqnarray}
Moreover, the spectral function is an essential ingredient to
compute the momentum distribution

\begin{equation}
n({\bf q})=\langle a_{\bf q}^{\dagger }a_{\bf q}\rangle
=\int_{-\infty }^{0} {\cal A}({\bf q},\omega ) d\omega \label{nq}
\end{equation}
and the single particle density matrix given by the Fourier transform
of the momentum distribution in real space

\begin{eqnarray}
\rho(\mathbf{r},{\mathbf{r}^\prime})=\langle a_{\mathbf{r}}^\dagger
a_{\mathbf{r}^\prime} \rangle = \frac{1}{N} \sum_{\bf q} e^{i{\bf
q}\cdot(\mathbf{r}-\mathbf{r}^\prime)} n({\bf q}), \label{rho0}
\end{eqnarray}
where $N$ indicates the number of lattice wells . The long distance
behavior of $\rho(\mathbf{r},{\mathbf{r}^\prime})$ as a function of
the relative distance approaches the condensate density $n_0$ which is
non zero in a SF and vanishes in the MI.
In the following we will calculate and discuss all these quantities.

A fundamental implication of broken symmetry for bosonic systems is
that the Goldstone modes are directly reflected in the single particle
spectrum. In other words, phonons which are related to modes of
density-density fluctuations (or two-particle Green function) also
show up as the poles in the single particle Green
function~\cite{griffin}. We study the behavior of the poles of the
Green's function, their strength, momentum and frequency dependence,
to extract information about the excitations of the system. In the two
extreme limits of deep MI and weakly interacting (Bogoliubov) SF, the
Green's function can be calculated analytically and from it, the
excitations frequencies and the momentum distribution $n({\bf q})$.

\subsection{Deep Mott regime}

In the deep Mott regime (zero tunneling), for $U(n-1)<\mu < Un$,
one finds

\begin{eqnarray}
G_{MI}({\bf q},\omega ) &=&\frac{1}{2\pi }\left[ \frac{n+1}{\omega
-(Un-\mu )}- \frac{n}{\omega -(U(n-1)-\mu )}\right]  \nonumber
\label{greenmott} \\
&& \\
\omega _{MI} ({\bf q})&=&\left\{
\begin{array}{ll}
Un-\mu  >0  &  \\
U(n-1)-\mu <0  &
\end{array}
\right.
\label{omegaMott} \\
n_{MI}({\bf q}) &=&n,\;\forall \;{\bf q}, \label{nmott}
\end{eqnarray}
where $n$ is the atomic occupation at each lattice site. The spectral
function consists of two $\delta$-functions, one at positive energy
$Un-\mu$ (relative to the chemical potential) required to add a
particle and one at negative energy $U(n-1)-\mu$, required to remove a
particle or add a hole to the MI, as seen in
Eq.~(\ref{omegaMott}). The spectral function ${\cal A}({\bf
  q},\omega)$ obtained using expression (\ref{greenmott}) trivially
satisfies the sum rule in Eq.~(\ref{sumrule}). The momentum
distribution, as defined in Eqs.~(\ref{nq},\ref{nmott}), is completely
flat, corresponding to vanishing site-to-site correlations, and
normalized to the total number of atoms in the lattice ($n$ times the
number of sites).  Correspondingly, the single particle density matrix
(\ref{rho0}), given by the Fourier transform of the momentum
distribution, shows strictly on-site correlations.

\subsection{Weakly interacting regime}

In the weakly interacting SF regime \cite{pitaevski-stringari}, we have

\begin{eqnarray}
G_{BG}({\bf q},\omega ) &=&\frac{1}{2\pi }\left[ \frac{|u_{\bf
q}|^{2}}{\omega -\omega _{\bf q}}-\frac{|v_{-{\bf q}}|^{2}}{\omega
+\omega _{-{\bf q}}}\right] , \label{greensf} \\ \omega _{BG}({\bf q})
&=&\pm \omega _{\pm {\bf q}},
\label{omegaBogo} \\
n_{BG}({\bf q}) &=& n_0 \delta_{{\bf q},0} + |v_{-{\bf q}}|^{2},
\label{nsf}
\end{eqnarray}
where $u_{{\bf q}}$ and $v_{{\bf q}}$ are the Bogoliubov amplitudes,
$\omega _{{\bf q}}$ is the Bogoliubov frequency at momentum ${\bf q}$
(see App.~\ref{app2} for details), and $n_0$ the condensate
density. In the weakly interacting SF regime, the sum rule in
Eq.~(\ref{sumrule}) is constrained by the Bogoliubov normalization
condition $|u_{{\bf q}}|^{2}-|v_{-{\bf q}}|^{2}=1$.  The excitation
energies are given by symmetric poles at positive and negative
frequencies corresponding to the energies of the Bogoliubov spectrum,
highlighting in particular phononic excitations at low momentum.

The momentum distribution, given by integrating the spectral function
over negative energies, has a singular contribution from the
condensate at zero momentum. The integral over all momenta different
from zero gives the number of non condensed atoms, contributing the
depletion from the condensate. In the regime where Bogoliubov theory
is valid, the depletion is negligible compared to the atoms in the
condensate.

\subsection{Progression from SF to MI}

The RPA formalism allows us to calculate the spectral function
with special emphasis on the strong correlation region near the
QPT. In the deep SF, we find phonon collective modes reflected in
the single particle spectrum. As $t/U$ is decreased, the spectral
weight is redistributed over a multi-mode structure composed by
coherent phonon excitations and gapped single-particle
excitations. When entering the MI phase at the QPT, the spectral
weight reorganises and is shared by only two gapped modes,
describing single particle excitations, one at positive and one at
negative energy. In the following, we will deiscuss this behaviour
more in detail.

\begin{center}
\begin{figure}[t!]
\includegraphics[width=0.92\linewidth]{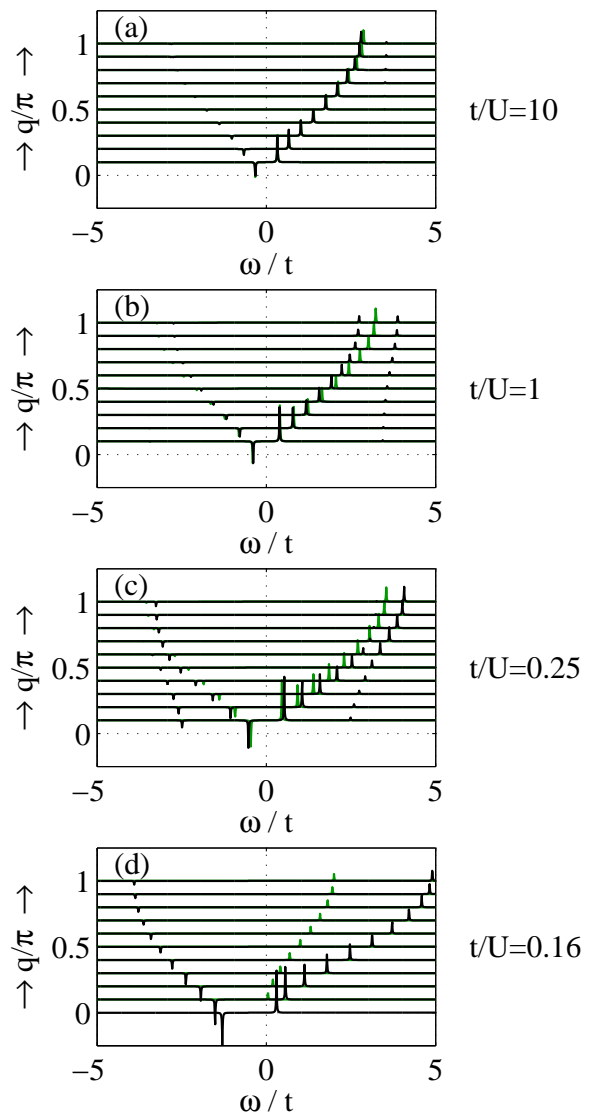}
\caption{(Color online) Spectral function ${\cal A}(q,\omega)$ as
a function of $\omega$ for various $q$. Results obtained by RPA
(black) and Bogoliubov theory (green). (a) Weakly interacting SF:
$t/U=10$ with $\approx 10$ bosons per site; the RPA calculation
agrees extremely well with the Bogoliubov theory; notice
indications of additional modes at higher $\omega$.  (b) $t/U=1$
with $\approx 1.8$ bosons per site. (c) Strongly interacting SF:
$t/U=0.25$ with $\approx 1.1$ bosons per site; stronger deviations
from Bogoliubov theory are present especially at larger $q$.
Additional modes are clearly visible in the spectrum. (d) Mott
insulating phase for $t/U=0.16$ and $1$ boson per site. In all
those figures $\mu/U=0.5$.} \label{green}
\end{figure}
\end{center}

We use the position of the poles of the Green's function to determine
the following results about the excitations of the system in the
different regimes:

(i) For a large number of particles per site ($\approx 100$) and weak
interactions ($t/U\approx 100$), we exactly recover the Bogoliubov
results. We point out that being able to describe the weakly
interacting regime starting from the BHM is not a trivial result,
because of the large number of basis states required (almost 150
states per site).

(ii) By increasing the interactions and decreasing the number of
particles per site, we observe small deviations between the spectrum
obtained by RPA and the Bogoliubov theory: additional modes appear at
higher frequencies, as shown in Fig.~\ref{green}(a) for $t/U=10$, in
contrast with Bogoliubov theory which predicts a single excitation
mode. While there are also differences in the dispersion of the sound
modes at large momenta, we find in general that the Bogoliubov
prediction turns out to be quite accurate in describing the low-$q$
part of the spectrum and the sound velocity, even in the case of
strong interactions (see Sect.~\ref{sect_c}).

\begin{center}
\begin{figure}[h!]
\includegraphics[width=0.85\linewidth]{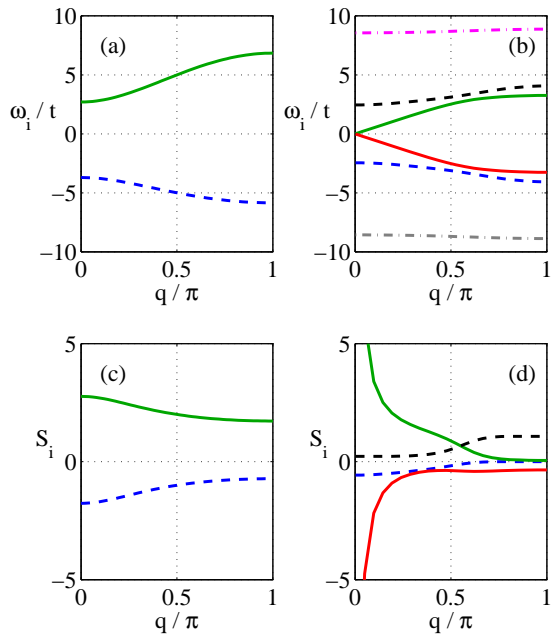}
\caption{(Color online) Dispersion (a,b) and strength (c,d) of
excitation modes at $\mu/U=0.5$: (a) and (c) are in the Mott
regime $t/U=0.1$; (b) and (d) are in the SF regime $t/U=0.25$. For
clarity, in (b) the 4th pair of resonances at $\omega\approx \pm
18.4$ is not shown and in (d) only the strength of the 4 modes at
lower energy is shown. Note that modes at positive (negative)
energy have positive (negative) strength.} \label{excitations}
\end{figure}
\end{center}

(iii) As $t/U$ becomes of order unity and the effect of strong
correlations grows, additional gapped modes in the SF phase are
clearly visible and grow in strength as seen in
Fig.~\ref{green}(b). The phonon modes are not sufficient to exhaust
the sum rule in Eq.~(\ref{sumrule}). In a strongly interacting SF (e.g
for $t/U<0.25$, as shown in Fig.~\ref{green}(c)), many modes (in the
cases we have considered, up to four at positive and four at negative
energy) have to be included in order to exhaust the sum rule in
Eq.~(\ref{sumrule}).  In the particular case of $\mu/U=0.5$ and
$t/U=0.25$, the full dispersion of the modes is shown in
Fig.~\ref{excitations}(b).

(iv) In the MI , only two excitation modes exist, as shown in
Figs.~\ref{green}(d) and \ref{excitations}(a).  The mode at positive
and the one at negative energy correspond respectively to the energy
needed to create a particle or a hole in the system. For any given
$t/U$, the difference of the excitation energies at $q=0$ exactly
coincides to the width of the mean-field Mott lobe at the same $t/U$.

\subsection{Strengths of the spectral function}

The progression of the modes from the strongly correlated SF into the
MI is better understood by calculating the strengths of the
excitations $S_{i}$, defined as follows:

\begin{eqnarray}
\mathcal{A}({\bf q},\omega)&=& \sum_i S_i
\delta(\omega-\omega_i). \label{AS}
\end{eqnarray}
Numerically, a small but finite imaginary part of the energy
regularizes the spectral function and provides an accurate fitting
procedure to determine the position of the poles and their
strength. We checked that the sum rule in Eq.~(\ref{sumrule}), which
using Eq.~(\ref{AS}) implies $\sum_i S_i =1$, was found to be
satisfied to better than few parts in $10^{-5}$ for all $t/U$.  We are
therefore confident that we have identified all the excitations which
contribute in a non-negligible way to the spectrum.

\begin{center}
\begin{figure}[h!]
\includegraphics[width=0.8 \linewidth]{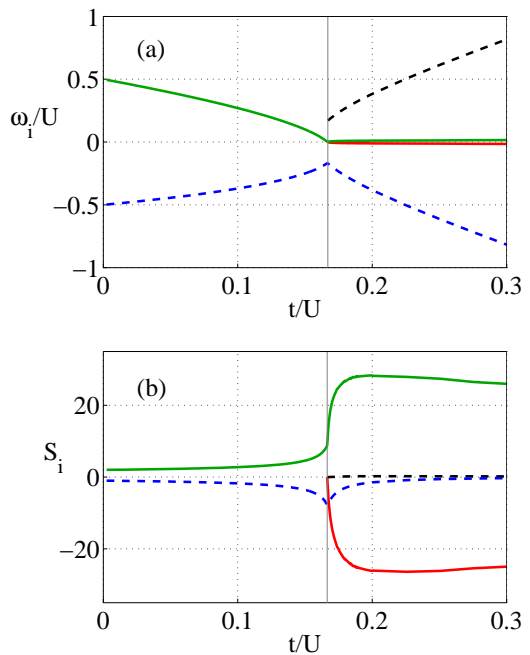}
\caption{(Color online) Energy (a) and strength (b) of the modes
at low $q=\pi/50$ as a function of $t/U$ for $\mu/U=0.5$. Notice
the presence of both phonons and gapped modes in the strongly
interacting SF and their evolution into two gapped modes into the
MI. One of the gapped modes in the MI evolves from the phonon mode
in the SF and the other one from a gapped mode in the SF. The thin
vertical line at $t/U\approx0.167$ indicates the QPT.}
\label{excitations2}
\end{figure}
\end{center}

\begin{center}
\begin{figure}[h!]
\includegraphics[width=0.8 \linewidth]{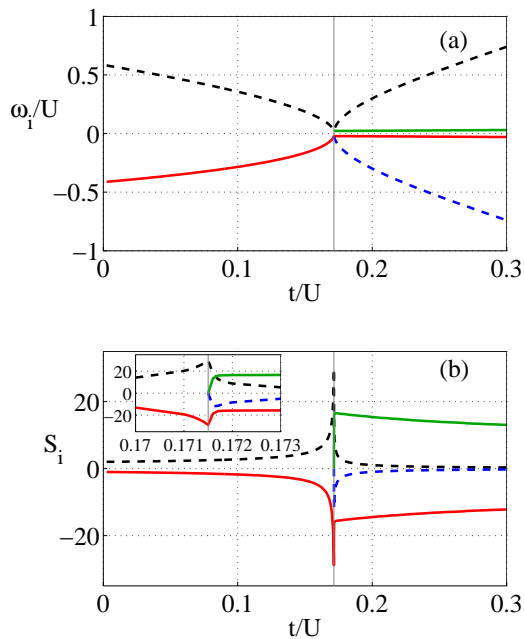}
\caption{(Color online) Energy (a) and strength (b) of the modes
at low $q=\pi/50$ as a function of $t/U$ at $\mu/U=\sqrt{2}-1$
corresponding to the tip of the lobe. The frequency of the lowest
four modes (two phononic and two gapped) in the SF vanish at the
QPT. The thin vertical line at $t/U\approx0.1716$ indicates the
QPT. In the inset, the zoom around the QPT of panel (b) is shown.}
\label{tip}
\end{figure}
\end{center}

In Figs.~\ref{excitations2},\ref{tip}, we plot the position of the
resonances and their strengths for a fixed value of $q=\pi/50$,
varying the parameter $t/U$ across the phase transition for fixed
chemical potential. As explained above, the many-mode spectrum in
the SF phase evolves into the two-mode excitation spectrum in the
MI.

The transition from the MI to the SF phase occurs when one of the two
Mott branches becomes gapless (see Fig.~\ref{excitations2}(a)). This
is the particle (hole) gapped mode in the MI depending on whether the
chemical potential $\mu$ is above (below) the tip of the lobe. This
Mott branch evolves into the phononic mode at positive (negative)
energy, while the other phononic mode arises at zero strength without
having a precursor in the Mott phase. The second Mott branch evolves
into a gapped superfluid mode, and symmetrical in energy a second
superfluid gapped mode arises with zero strength without having a
precursor in the Mott phase (see Fig.~\ref{excitations2}(b)).

The behavior at the tip of the lobe is quite interesting. In that
case both Mott gapped modes become simultaneously gapless at the
QPT (see Fig.~\ref{tip}(a)). From them the lowest four modes in
the SF arise, two of them becoming phononic modes and two of them
becoming gapped modes when moving away from the transition.
However, within our approach, we find a similar behavior to the one
described above, namely that one of the Mott modes evolves into a
SF phononic mode, while the second one evolves into a SF gapped
mode (see Figs.~\ref{tip}(b)).

A similar result was found by Huber and collaborators
~\cite{huber_prb2007} using an effective 3-state approximation and
mapping onto a spin-1 Hamiltonian. They further suggested that
measurements of the dynamical structure factor using Bragg
spectroscopy and lattice modulations could be an effective way to
investigate the different modes.

The problem was also investigated by Sengupta and Dupuis
~\cite{sengupta_pra2005} in the strong coupling regime by deriving an
effective action using Hubbard Stratonovich transformations. By
expanding the action to quadratic order in the fluctuations they found
gapped excitations in the MI and gapless Goldstone modes in the
SF. They found two additional gapped modes in the SF, which present a
similar behavior to the one discussed in this paper.

\subsection{Density of states}

\begin{center}
\begin{figure}[t]
\includegraphics[width=0.9\linewidth]{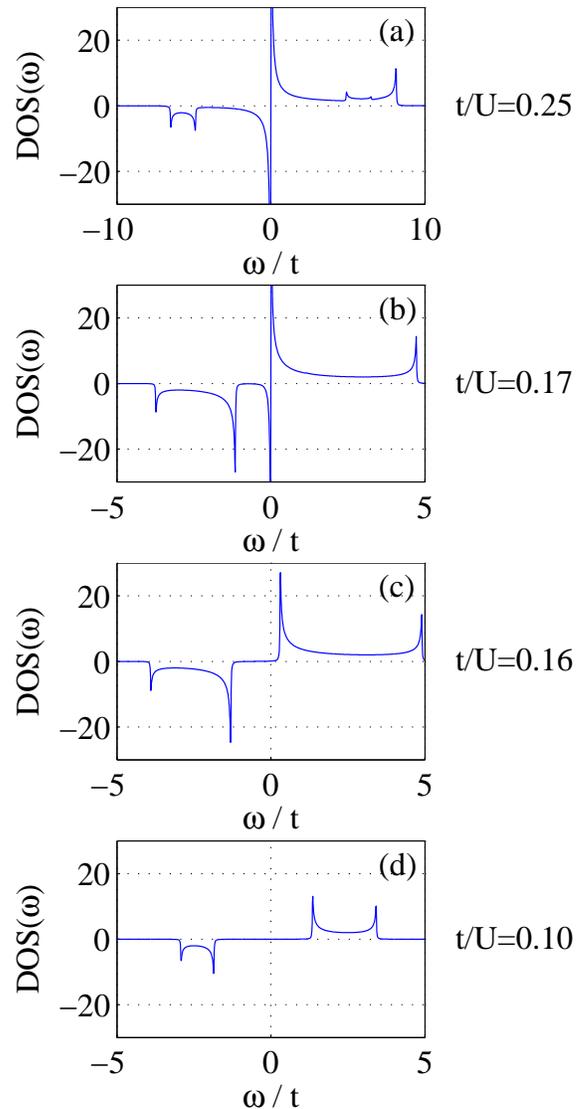}
\caption{Density of states for several values of $t/U$. (a)
$t/U=0.25$, SF regime; (b) $t/U=0.17$, SF regime, very close to the
QPT; (c) $t/U=0.16$ MI regime, very close to the QPT; (d) $t/U=0.1$ MI
regime. All those results are for $\mu/U=0.5$.}
\label{fig_dos}
\end{figure}
\end{center}

A further quantity that one can use to characterise the
excitations of the system is the density of states defined in
Eq.~(\ref{dos}) \cite{sengupta_pra2005, ohashi_pra2006}. We
calculate it across the QPT for a 1D system \cite{footnote}, as
shown in Fig.~\ref{fig_dos}. In Fig.~\ref{fig_dos}(a) one can
recognize the multi-mode structure of a strongly correlated SF
through a clearly enhanced DOS in the energy range of the
corresponding excitation branch. In particular, we can see here
two phononic branches and two gapped ones. When approaching the
QPT ($t/U=0.17$ and Fig.~\ref{fig_dos}(b)), we encounter a similar
structure, where the width of the gapped branch at negative energy
is increased, while the gapped branch at positive energy is not
visible on the scale of this picture (although existing), since
its strength goes to zero at the QPT. As expected, in the Mott
regime, the DOS is different from zero only in the energy range of
the two gapped excitation branches, one at negative and one at
positive energy (Figs.~\ref{fig_dos}(c,d)). As observed in
Fig.~\ref{fig_dos}(c), the branch at positive energy extends
almost to $\omega=0$, indicating the disappearance of the gap at
the QPT.

\section{Sound velocity} \label{sect_c}

The presence of phononic modes in the excitation spectrum is an
important signature of superfluidity. These modes disappear in the
Mott phase, where sound cannot propagate because of a gap in the
spectrum. In this section, we discuss the evolution of the sound
velocity in the strongly correlated SF phase as the SF-MI transition
is approached.

The sound velocity is related to the compressibility $\kappa$ and the
effective mass $m^\ast$ \cite{zaremba, kraemer_epjd, menotti_pra2004}
through the relation

\begin{eqnarray}
c = \frac{1}{\sqrt{\kappa m^*}}=\sqrt{\frac{\rho_s}{\rho\kappa}m},
\label{kompr}
\end{eqnarray}
where $\kappa^{-1}=\rho (\partial \mu / \partial \rho)$ and the SF
fraction $\rho_s/\rho=m/m^*$. We calculate the sound velocity $c$ from
the slope of the gapless mode in the RPA spectrum

\begin{eqnarray}
\lim_{|{\bf q}|\ \to 0} \omega({\bf q}) = c |{\bf q}|,
\label{dispersion}
\end{eqnarray}
and compare it with the one obtained with the compressibility relation
in Eq.~(\ref{kompr}). We find perfect agreement between the values for
the sound velocity extracted by the two different methods. It is
important to note that the method using the compressibility relation
in Eq.~(\ref{kompr}) only requires the knowledge of the mean-field
solution, which provides the equation of state $\mu(\rho)$ and the SF
density $\rho_s=|\varphi|^2$.

\begin{center}
\begin{figure}[h!]
\includegraphics[width=0.95\linewidth]{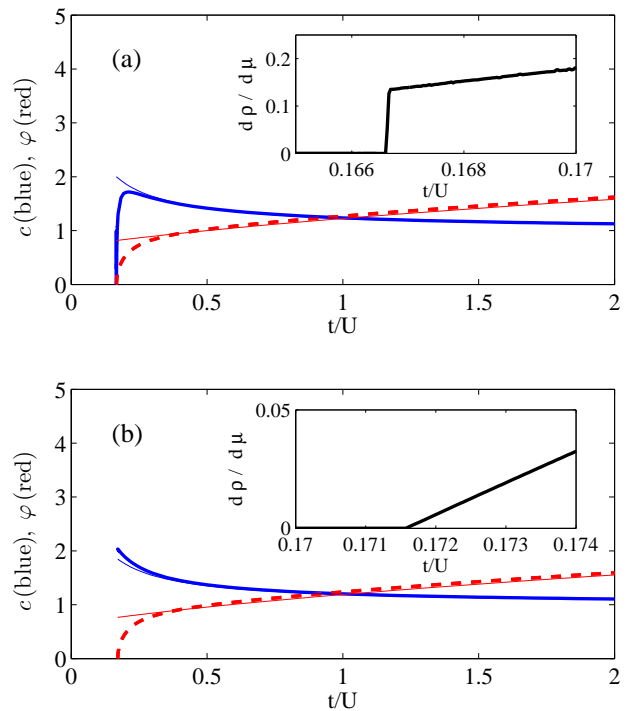}
\caption{(Color online) Sound velocity $c$ (blue full line)
extracted from the slope of the dispersion of the phonon modes and
from Eq.~\ref{kompr}; order parameter $\varphi$ (red dashed line).
Also shown for comparison are the sound velocity and the order
parameter obtained from Bogoliubov theory (thin lines). In the
inset $d\rho/d\mu$ is shown. (a) $\mu/U=0.5$; (b)
$\mu/U=\sqrt{2}-1$, corresponding to the tip of the lobe.}
\label{sound}
\end{figure}
\end{center}

We find that when the SF-MI transition, tuned by $t/U$ and $\mu/U$, is
approached at a generic point away from the tip of the lobe, the sound
velocity vanishes, as shown in Fig.~\ref{sound}(a). This is due to the
fact that at the transition the compressibility remains finite, but
the SF density vanishes. Instead, the tip of the lobe where the phase
transition happens at constant density and ${\partial \rho}/{\partial
\mu}=0$ is a special point: there, a perfect compensation between the
divergent inverse compressibility and the vanishing SF density takes
place, which results in a finite sound velocity as seen in
Fig.~\ref{sound}(b).

We complete our analysis by comparing the sound velocity calculated
above with the results of Bogoliubov theory, which for a tunneling
parameter $t$, on-site interaction $U$ and coordination number $z$,
predicts the value

\begin{eqnarray}
c_{BG}= \sqrt{ \frac{2 t}{z} U |\varphi|^2}, \label{cbog}
\end{eqnarray}
as explained in detail in App.~\ref{app2}. The Bogoliubov
predictions are remarkably good in a wide range of parameters and
fail only in the close proximity of the phase transition (see thin
lines in Fig.~\ref{sound}), since Bogoliubov theory does not
account correctly for the vanishing of the order parameter at that
point.

\vspace{0.5cm}

\section{Momentum distribution and spatial correlations}
\label{nqrhorr}

From Eq.~(\ref{nq}), the momentum distribution $n({\bf q})$ is
obtained by integrating the spectral function over negative
energies. It is a quantity of primary importance in cold atom
experiments, as it is directly accessible by imaging the cloud after
expansion~\cite{bloch_nature2002,spielman_prl2007,diener_prl2007}.  We
have considered a two-dimensional system, which allows for the
existence of Bose-Einstein condensation with long range order in the
SF regime.

\begin{center}
\begin{figure}[h!]
\includegraphics[width=0.95\linewidth]{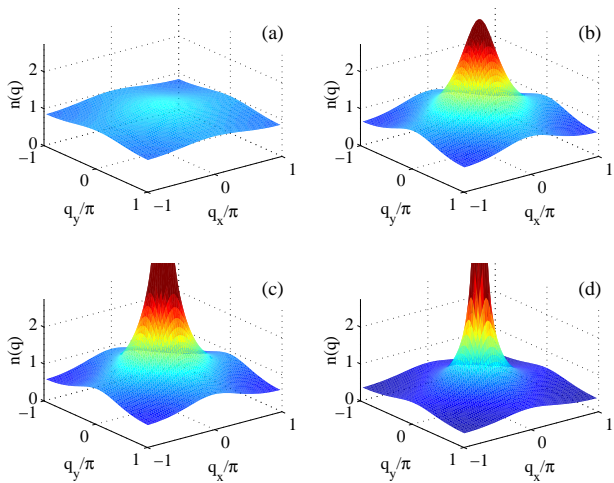}
\caption{(Color online) Momentum distribution $n(q)$ for a 2D system
for $\mu/U=0.5$ and (a) $t/U=0.05$ deep in the MI; (b) $t/U=0.15$ in
the MI but closer to the QPT; (c) $t/U=0.175$ in the SF phase close to
the QPT; (d) $t/U=0.25$ in the SF phase but further from the QPT.}
\label{fig_nq}
\end{figure}
\end{center}

The 2D momentum distribution is shown in Fig.~\ref{fig_nq} for
different values of the parameter $t/U$. We have checked that we can
reproduce the momentum distribution in the two extreme cases of deep
MI (see Eq.~(\ref{nmott})) and weakly interacting SF (see
Eq.~(\ref{nsf})).

In the Mott phase with non vanishing tunneling, we find that the
momentum distribution presents a modulation showing up as
interference peaks in the expansion pictures \cite{bloch_prl2005,
spielman_prl2007}. When the QPT is reached, a strong peak develops
at $q=0$ corresponding to the condensate. However, in a SF close
to the phase transition the background momentum distribution (at
$q \neq 0$) is large indicating a strong depletion from the
condensate due to interactions.

The momentum distribution obtained with the RPA method happens to be
not correctly normalized to the total number of atoms. We attribute
this feature to the fact that fluctuations are not self-consistently
included in the ground state (see discussion in App.~\ref{app3}).

Starting from the momentum distribution, one can obtain the single
particle density matrix $\rho(\mathbf{r},{\mathbf{r}^\prime})$, which
contains direct information about the spatial correlations present in
the system. In the SF phase, the system is characterized by
off-diagonal long range order, and at long distances, the single
particle density matrix approaches a constant value equal to the
square of the order parameter, or the condensate density $n_0$

\begin{equation}
\rho({\bf r},{\bf r^\prime)}\equiv\langle a^\dagger ({\bf r})
a(0)\rangle \rightarrow \varphi^2 = n_0 , \label{odlro}
\end{equation}
which is non-zero in a SF. This quantity is linked through Fourier
transform to the $\delta$-function at ${\bf q}=0$ which appears in the
momentum distribution.

\begin{center}
\begin{figure}[h!]
\includegraphics[width=0.95\linewidth]{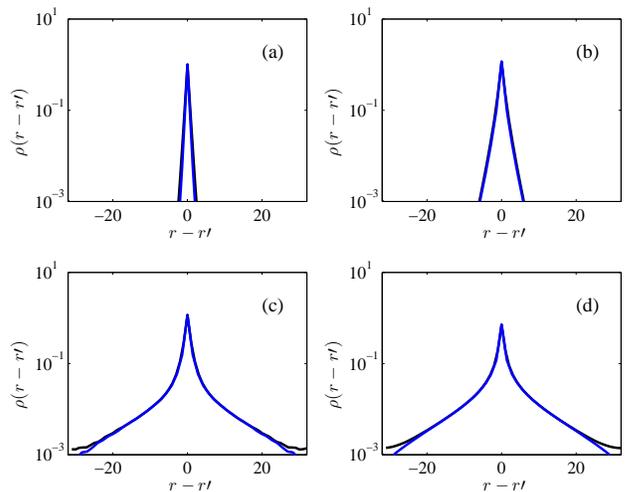}
\caption{(Color online) Density matrix $\rho(r,r^\prime)$ as a
function of the relative distance $r-r^\prime$ for a 2D system (black
line for a cut in the center of the trap and blue line along the
diagonal). (a) $t/U=0.05$ deep in the MI showing only nearest neighbor
correlations; (b) $t/U=0.15$ in the MI but closer to the QPT showing
an increase in the scale of the short range correlations; (c)
$t/U=0.175$ in the SF phase close to the QPT showing long range order;
(d) $t/U=0.25$ in the SF phase but further from the QPT. Note that the
asymptotic value for large $r-r^\prime$ has been subtracted. }
\label{rhorr}
\end{figure}
\end{center}

The single particle density matrix in Fig.~\ref{rhorr} shows a marked
transition from the MI phase to the SF, and corresponds to a change
from an exponential decay of the correlations to a (quasi) long range
order.  In the MI, the single particle density matrix shows that the
correlations decay over a finite length scale, which decreases by
decreasing the tunneling and moving deeper into the Mott lobe. From
those results, we have direct access to the length scale of the
correlations in the insulating regime.

The condensate fraction can be in principle extracted from the
momentum distribution, by subtracting the depletion $\sum_{{\bf q}\neq
0} n({\bf q})$ from the total density $n$, or equivalently by looking
at the asymptotic value at large distances of the single-particle
density matrix $\rho({\bf r},{\bf r}')$.  However, unfortunately we
find that the present application of RPA gives a violation of the
total density sum rule and $\sum_{{\bf q}\neq 0} n({\bf q})>n$ (see
also discussion in App.~\ref{app3}), so that neither the momentum
distribution nor the single-particle density matrix turn out to be
useful quantities to extract the condensate density. This problem
arises because within the present theoretical description, we have not
included the feedback of the collective modes and other excitations
into the mean field ground state. Possible solutions to this problems
will be topic of further research~\cite{diener-randeria2007}.

However, our analysis of the excitation modes and their strength
allows us to extract the condensate density $n_0$ directly, using
our knowledge of the sound velocity (from the slope of the
phononic mode at small momenta), combined with the knowledge of
the strengths of the spectral function and the compressibility.
Starting from the relation

\begin{eqnarray}
\mathcal{A}(q,\omega)=\frac{n_0}{\rho_s}mc^2\delta(\omega^2-c^2q^2),
\label{condfract}
\end{eqnarray}
and assuming that the condensate and SF densities are equal (at least
within mean field theory), we get

\begin{eqnarray}
n_0=\frac{(2qS_{ph})^2}{m^* d\mu / d\rho}=\frac{(4\pi q
S_{ph})^2\;t}{d\mu /d\rho}. \label{n0vsSph}
\end{eqnarray}
where $S_{ph}$ is the strength of the phononic modes. We have used
$t=\hbar^2/(m^* d^2)$ (with $\hbar=1/2\pi$ and lattice spacing $d=1$)
to relate the effective mass and the tunneling parameter in the BHM.

\begin{center}
\begin{figure}[h!]
\includegraphics[width=0.65\linewidth]{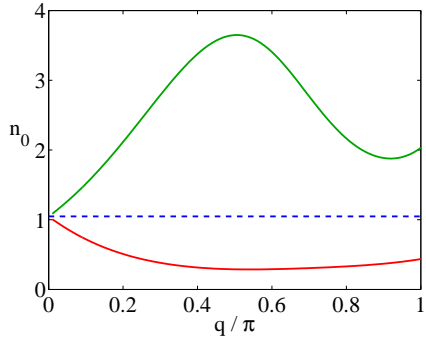}
\caption{(Color online) Condensate fraction obtained from the
strength of the poles of the spectral function, as in
Eq.~(\ref{n0vsSph}), for the phonon mode at positive (upper green
line) and negative (lower red line) frequency. As $q\rightarrow 0$
both functions approach the condensate fraction $n_0$ obtained
with MFT (dashed blue line). In this figure $\mu/U=0.5$ and
$t/U=0.5$.} \label{Aqw}
\end{figure}
\end{center}

The limiting behavior of the RHS of Eq.~(\ref{n0vsSph}) for very small
$q$ approaches the condensate density which coincides with the
predictions of mean-field theory ($|\varphi|^2$), as shown in
Fig.~\ref{Aqw}.

\section{Inhomogeneous system: Optical lattices in an external trapping
potential} \label{sect_inhom}

We extend the RPA formalism to real space and include a spatially
inhomogeneous potential, which is taken into account in the BHM
through a site-dependent chemical potential. The self consistent MF
solution produces alternating shells of insulating and superfluid
phases moving out from the center to the edge of the
trap~\cite{jaksch_prl1998, batrouni_prl2002, bloch_prl2006}.

In the inhomogeneous system, the derivation of the equation for the
Green's function is the same outlined in App.~\ref{app1}, with the
essential caution that the on-site energies $\epsilon_\alpha^i$ and
the tunneling coefficients ${\tilde T}^{i k }_{\alpha^{\prime }\alpha
\gamma \gamma^{\prime }}$ depends on position and must be calculated
respectively for each site and for each pair of neighboring sites.

\begin{center}
\begin{figure}[h!]
\includegraphics[width=1\linewidth]{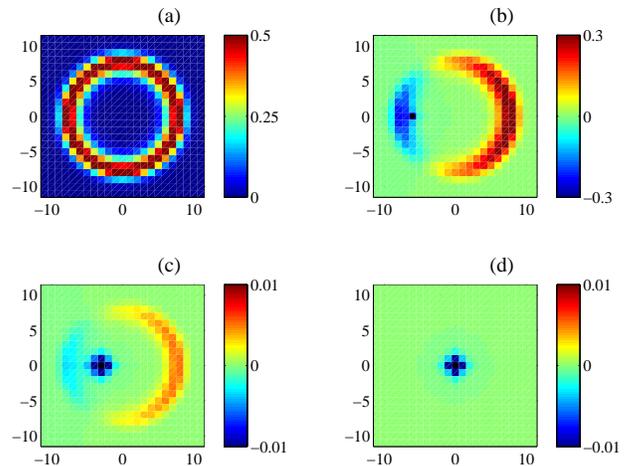}
\caption{(Color online) Inhomogeneous system: (a) order parameter in
  the trap: It is zero in the central MI core and finite and large in
  the surrounding SF ring. The other panels show $G({\bf r},{\bf
    r^\prime})$ for a given value of the energy as a function of ${\bf
    r^\prime}$ for a fixed ${\bf r}$ (black dot): (b) ${\bf r}=(-6,0)$
  in the SF ring; (c) ${\bf r}=(-3,0)$ at the interface between the MI
  core and the SF ring; (d) ${\bf r}=(0,0)$ in the center of the MI
  core.}
\label{Gtrap}
\end{figure}
\end{center}

In the presence of an external trapping potential, the density and
order parameter become non-uniform as shown in
Fig.~\ref{Gtrap}(a). With our specific choice of parameters, one finds
a central Mott core at density $n=1$, surrounded by a ring of
superfluid. The sequence of panels (b) to (d) shows $G({\bf r},{\bf
r'},\omega)$ as a function of ${\bf r'}$ for fixed ${\bf r}$ and
$\omega$, which roughly speaking represent the effect of perturbing
the system at different points: perturbations in the SF regions
produce a large effect all along the SF ring. As the perturbation
moves to regions with lower order parameter near the SF-Mott interface
its effect gets reduced and finally perturbations in Mott-like regions
decrease exponentially and produce negligible effects.

The results shown in Fig.~\ref{Gtrap} are obtained for a given
value of the energy $\omega$. At different energies the structure
remains the same, but the period of the oscillations along the
ring changes. By integrating $G({\bf r},{\bf r'},\omega)$ over
$\omega$, one gets the equal-time correlation function $G({\bf
r},{\bf r'})$. This quantity would maintain the ring-like behavior
shown Fig.~\ref{Gtrap} and hence be qualitatively similar to the
one calculated by Wessel and collaborators \cite{wessel_pra04}.

Before searching for quantitative results in the non uniform system,
one should ponder on the consequences of the problem in the
normalization of the momentum distribution discussed in
Sect.~\ref{nqrhorr}, which might affect the extension of the RPA method
to trapped systems. While this extension is technically simple,
although it might become computationally quite expensive, its validity
is to be questioned due to the co-existence in the same system of
different phases (MI and SF), where, as we have explained above, RPA
introduces different normalization factors.  For this reason, while we
believe one can get some insights on the correlations in the system,
those pieces of information can be trusted only at the qualitative
level.

\section{Concluding remarks: RPA vs mean-field}
\label{conclusions}

In this work, we have studied the excitations and the spatial
correlations of the BHM in the RPA approximation. It is interesting to
compare the information that is obtained between the mean field
approximation and the RPA which includes a certain class of
fluctuations:

(1) The prediction of the SF to MI transition in the $[\mu/U,t/U]$
plane calculated from the vanishing of the order parameter
$\varphi$ at the mean-field level and by the disappearance of the
gapless excitation mode within RPA lead to exactly the same result
for the boundaries of the insulating lobes.

(2) At the mean field level, one can extract information about the
sound velocity using the compressibility-effective mass relation,
while at the RPA level the sound velocity is given by the slope of the
phonon mode. The two methods again lead to exactly the same result.

(3) The condensate fraction is $n_0=|\varphi|^2$ at the mean-field
level, whereas in the RPA treatment, it is extracted from the small
$q$ behavior of the spectral function and using the mean-field
compressibility. Again the two methods lead to exactly the same
result.

If the fluctuations around the mean field state were included
self-consistently, they would renormalize the energy and order
parameter. We then expect the SF state to be susceptible to
fluctuations and the critical $t/U$ to be shifted to a higher value
than the one obtained by the MF theory.


The RPA method further gives information which is not included in the
mean-field treatment. These include (i) the excitation spectra; (ii)
their strengths; (iii) the existence of new gapped modes in the
strongly interacting SF phase; (iv) the momentum distribution and
spatial correlations in the system.


As an open question, we are left with the role of quantum
fluctuations in the vicinity of the QPT, which as we discussed may
have additional effects on physical quantities like the e.g.
momentum distribution or the condensate fraction. As explained in
App.~\ref{app3}, it is to be expected that in our approach the
momentum distribution is not normalized. In the Mott limit, a
clear deviation from the normalization to the total number of
atoms can be calculated analytically. Analogously, in the
Bogoliubov regime, exactly recovered in RPA in the dilute limit,
where one assumes $n=|\varphi|^2$, the normalization is larger
that the total number of atoms once the depletion (given by the
integral over all momenta different from zero of $|v_q|^2$) is
added. However, while in the deep MI and SF regimes, the change in
the normalization is just a small perturbation, close to the phase
transition it is a striking effect. We attribute this to the fact
that we perform RPA on the mean-field ground state, without taking
the effect of RPA self-consistently into account. To this same
reason, we attribute the existence of predictions (1) to (3) which
are equal in the mean-field and RPA treatments.

\begin{acknowledgments}
C.M. acknowledges financial support by the EU through an EIF
Marie-Curie Action. We thank R.B.~Diener, P.~Pedri, M.~Randeria, and
S.~Stringari for helpful discussions.
\end{acknowledgments}

\appendix

\section{Green's function formalisms in the RPA approximation}
\label{app1}

In this appendix we recall the main steps of the derivation of the
Green's function formalism in the random phase approximation (RPA)
\cite{haley,zubarev}.

RPA includes some fluctuations around the mean-field solution, which
allows us to describe the excitations of the system.  However, as
explained in Sect.~\ref{conclusions}, these fluctuations are not
included self consistently, allowing feedback into the mean field
ground state. Also ignored are quantum fluctuations of the order
parameter which are especially important close to the QPT (see also
discussion in App.~\ref{app3}).

\emph{Mean-field decoupling:}

Substituting $a=\varphi + {\tilde a}, \;\; a^{\dag} =\varphi + {\tilde
a}^{\dag}$, in Eq.~(\ref{bhm}), we obtain without any approximation

\begin{eqnarray}
H = \sum_i \left[ \frac{U}{2} n_i(n_i-1) - \mu n_i - t
\varphi(a^{\dag}_i+a_i) + t \varphi^2 \right] + \nonumber \\
-\frac{t}{2z} \sum_{\langle i j \rangle} \left( {\tilde a}^{\dag}_i
{\tilde a}_j + {\tilde a}_i {\tilde a}^{\dag}_j \right). \hspace{1cm}
\label{hamil2}
\end{eqnarray}
The Hamiltonian $H$ is thus rewritten as a sum of on-site Hamiltonians
$H^{MF}_i$ (indicated by the term in the square bracket which
includes hopping at the mean field level), plus an inter-site hopping
term, which is assumed to be small.

\emph{Random Phase Approximation:}

In the basis given by the complete and orthonormal set of on-site
eigenstates $| i \alpha \rangle$ of the on-site Hamiltonians
$H^{MF}_i$, the Hamiltonian in Eq.~(\ref{hamil2}) takes the form

\begin{eqnarray}
H = \sum_{i \alpha} \epsilon_{\alpha}^i L^i_{\alpha \alpha}
-\frac{t}{2z} \sum_{ \langle i j\rangle \alpha \alpha^{\prime }\beta
  \beta^{\prime }} {\tilde T}^{ij}_{\alpha \alpha^{\prime }\beta
  \beta^{\prime }} L^i_{\alpha \alpha^{\prime }} L^j_{\beta
  \beta^{\prime }} ,
\label{hamil3}
\end{eqnarray}
where we have defined

\begin{eqnarray}
L^i_{\alpha \alpha^{\prime }}&=&|i \alpha\rangle \langle i \alpha^{\prime }|,
\nonumber \\
{\tilde T}^{ij}_{\alpha \alpha^{\prime }\beta \beta^{\prime }} &\equiv&
\langle i \alpha|{\tilde a}^{\dag}_i |i \alpha^{\prime }\rangle \langle j
\beta | {\tilde a}_j |j \beta^{\prime }\rangle +  \nonumber \\
&&+ \langle i \alpha|{\tilde a}_i |i \alpha^{\prime }\rangle \langle j \beta
|{\tilde a}^{\dag}_j | j \beta^{\prime }\rangle .
\label{T}
\end{eqnarray}
For any pair of single particle operators $A$ and $B$, the retarded
($\eta=+ 1 $) or advanced ($\eta=- 1$) Green's function, defined as

\begin{eqnarray}
G_{r,a}(\tau)=-i \eta \theta( \eta\tau) \langle A(\tau) B(0) -
B(0) A(\tau) \rangle ,
\label{gtau}
\end{eqnarray}
can be written in the on-site eigenbasis as

\begin{eqnarray}
G(\tau)=\sum_{\alpha\alpha^{\prime }\beta\beta^{\prime }} \langle
i \alpha| A |i \alpha^{\prime }\rangle \langle j \beta| B |j \beta
\rangle G^{ij}_{\alpha \alpha^{\prime }\beta \beta^{\prime }}
(\tau), \label{decomposition}
\end{eqnarray}
where $G^{ij}_{\alpha \alpha^{\prime }\beta \beta^{\prime }} (\tau) =
\langle\langle L^i_{\alpha \alpha^{\prime }}(\tau); L^j_{\beta
\beta^{\prime }} \rangle\rangle$.

In the energy domain, the Green's function is defined as

\begin{eqnarray}
G(\omega) &\equiv& \langle\langle A ; B
\rangle\rangle_{\omega_\pm} = \int_{-\infty}^{\infty}
\frac{dt}{2\pi} G(\tau)_{r,a} e^{i\omega_{\pm} \tau}.
\label{gomega}
\end{eqnarray}
A relation analogous to Eq.~(\ref{decomposition}) also holds for the
energy resolved Green's functions.

In the uniform system at vanishing tunneling (deep Mott regime), $H =
\sum_{i \alpha} \epsilon_{\alpha}^i L^i_{\alpha \alpha}$ is exactly
diagonalized by the on-site eigenstates (Fock basis). The Green's
function $G(\omega)$ can be calculated exactly to be

\begin{eqnarray}
G_{MI}^{t=0}(\omega)= \frac{1}{2\pi} \left[
\frac{n+1}{\omega-(E_{n+1}-E_n)}- \frac{n}{\omega+(E_{n-1}-E_n)}
\right] , \nonumber \\
\label{gmott_t0}
\end{eqnarray}
with $E_n=-\mu n+(U/2) n(n-1)$.

When the nearest neighbor tunneling plays a role, the commutator with
the transition operators $L^i_{\alpha\alpha^{\prime }}$ with the
tunneling part of the Hamiltonian produces a coupling to three
operators Green's functions.  Following the prescriptions of RPA of
replacing the average of a product with product of averages, one
obtains

\begin{eqnarray}
&&\left(E- \epsilon_{\alpha^{\prime }}^i + \epsilon_{\alpha}^i
\right) G^{ij}_{\alpha \alpha^{\prime }\beta \beta^{\prime }}(E) =
\label{afterRPAij}
\\
&&=\frac{1}{2\pi} (\langle L^i_{\alpha \alpha} \rangle-\langle
L^i_{\alpha^{\prime }\alpha^{\prime }} \rangle) \delta_{\alpha \beta^{\prime
}} \delta_{\alpha^{\prime }\beta} \delta_{ij} +  \nonumber \\
& - & \frac{t}{z} \left(\langle L^i_{\alpha \alpha} \rangle-\langle
L^i_{\alpha^{\prime }\alpha^{\prime }} \rangle \right) \sum_{ \langle k
\rangle_i \gamma \gamma^{\prime }} {\tilde T}^{i k }_{\alpha^{\prime }\alpha
\gamma \gamma^{\prime }} G^{kj}_{\gamma \gamma^{\prime }\beta \beta^{\prime
}} (E) .  \nonumber
\end{eqnarray}
At zero temperature results, $\langle L^i_{\alpha \alpha} \rangle$ are
equal to $1$ for the ground state and vanish otherwise.

For nearest-neighbor hopping and for a uniform system, where
$\epsilon_\alpha^i$, $\langle L^i_{\alpha \alpha} \rangle$, and
${\tilde T}^{ik}_{\alpha^{\prime }\alpha \gamma \gamma^{\prime }}$ are
site-independent (see Eq.~(\ref{T})), the same equation in momentum
space takes the form

\begin{eqnarray}
&&\left( E-\epsilon_{\alpha^{\prime }} + \epsilon_{\alpha} \right)
G_{\alpha \alpha^{\prime }\beta \beta^{\prime }} (E,{\bf q})=
\label{EqE} \\
&&= \frac{1}{2\pi} (\langle L_{\alpha \alpha}
\rangle-\langle L_{\alpha^{\prime }\alpha^{\prime }} \rangle)
\delta_{\alpha \beta^{\prime }} \delta_{\alpha^{\prime }\beta} +
\nonumber \\
&&+ \epsilon({\bf q}) \left(\langle L_{\alpha \alpha}
\rangle - \langle L_{\alpha^{\prime }\alpha^{\prime }} \rangle\right)
\sum_{\gamma \gamma^{\prime }} {\tilde T}_{\alpha^{\prime }\alpha
\gamma \gamma^{\prime }} G_{\gamma \gamma^{\prime }\beta \beta^{\prime
}} (E,{\bf q}) . \nonumber
\end{eqnarray}
where $\epsilon({\bf q})=-(2t/z) \sum_i\cos(q_i)$, with $i$
running over the dimensionality of the system and $z$ being the
number of nearest neighbors. In practice, for each value of the
energy $E$ and momentum ${\bf q}$, the solution of Eq.~(\ref{EqE})
amounts to inverting a $2(N_s-1) \times 2(N_s-1)$ matrix, where
$N_s$ is the dimension of the number state basis considered.

The solution can also be found analytically to be \cite{ohashi_pra2006}

\begin{eqnarray}
G({\bf q},\omega)=\frac{1}{2\pi} \frac{\Pi({\bf
q},\omega)}{1-\epsilon({\bf q}) \Pi({\bf q},\omega)},
\label{an_sol}
\end{eqnarray}
where
\begin{eqnarray}
\Pi({\bf q},\omega) &=& A_{11}+ \frac{A_{12}A_{21}
\epsilon(q)}{1-\epsilon({\bf q})A_{22}}, \\
A_{11} &=&
\sum_{\alpha} \frac{y_{0\alpha} y^\dag_{\alpha
0}}{\omega-\Delta_\alpha} - \frac{y_{0\alpha}^\dag y_{\alpha
0}}{\omega+\Delta_\alpha},
\label{a11_gen}\\
A_{22} &=& A_{11}^\dag, \\ A_{12} &=& \sum_{\alpha}
\frac{y_{0\alpha} y_{\alpha 0}}{\omega-\Delta_\alpha} -
\frac{y_{0\alpha} y_{\alpha 0}}{\omega+\Delta_\alpha}, \\ A_{21} &=&
A_{12}^\dag,
\end{eqnarray}
with $|0\rangle$ indicating the ground state, $y^{\dag}_{\alpha
0}=\langle \alpha |a^{\dag}|0\rangle$ (and analogously the other
terms), and $\Delta_\alpha=E_\alpha-E_0$. This equation can be easily
evaluated numerically once the mean-field ground state wave function
is known.  This can be done analytically in the MI phase (see
App.~\ref{app3}), but has to be done numerically in the SF phase.

In principle, it is possible to apply this formalisms also in the non
uniform case. One should then start from Eq.~(\ref{afterRPAij})
considering that both $\epsilon^i_{\alpha}$ and ${\tilde T}^{i k
}_{\alpha^{\prime }\alpha \gamma \gamma^{\prime }}$ generally depend
on position.  Hence, in that case, for each value of the energy the
solution amounts to inverting a $2(N_s-1)N \times 2(N_s-1)N$ matrix,
where $N$ is the number of lattice wells considered.

\section{Bogoliubov theory for the BHM}
\label{app2}

In this appendix we will present the details of the Bogoliubov
treatment for the BHM. The results are expected to be valid in the
weakly interacting SF regime, and were compared to the results of the
RPA approximation in Sect.~\ref{sect_c}.

We start from the BHM

\begin{eqnarray}
H=\sum_i \frac{U}{2} a^{\dag}_i a^{\dag}_i a_i a_i - \sum_i \mu n_i
-\frac{t}{2z} \sum_{\langle i j\rangle} \left( a^{\dag}_i a_j + a_i
a^{\dag}_j \right), \nonumber \\
\end{eqnarray}
and define, as before, the fluctuation operators subtracting from the
operators $a$ and $a^{\dag}$ their mean-value

\begin{eqnarray}
{\tilde a}=a - \varphi , \,\,\, {\tilde a}^{\dag}=a^{\dag}
-\varphi^*.
\end{eqnarray}
For a uniform system, one gets

\begin{eqnarray}
H &=& \sum_i \left[ \frac{U}{2} |\varphi|^4 - \mu |\varphi|^2 - t
|\varphi|^2 \right]
+ \\
&+& \sum_i \left[ {\tilde a}_i^{\dag} \left( U |\varphi|^2 -\mu -t
\right) \varphi
+ h.c. \right]  \nonumber \\
&+& \sum_i \left[ \frac{U}{2} \left( \varphi^2 {\tilde
a}_i^{\dag2} + 4 |\varphi|^2 {\tilde a}_i^{\dag} {\tilde a}_i +
\varphi^{*2} {\tilde a}_i^{2}
\right) + \right.  \nonumber \\
&& \left. \hspace{0.5cm}- \mu {\tilde a}_i^{\dag} {\tilde a}_i
-\frac{t}{2z} \sum_{\langle j\rangle_i} \left( {\tilde a}_i^{\dag}
{\tilde a}_j + {\tilde a}_j^{\dag}{\tilde a}_i \right) \right] +  \nonumber \\
&+& \mathrm{3rd + 4th \,\, order \;in\;} {\tilde a}\; {\mathrm
and\;} {\tilde a}^{\dag}. \nonumber
\end{eqnarray}
To minimize the energy one has to set to zero the first order, leading
to the discretized version of the Gross-Pitaevskii equation for the
uniform system

\begin{eqnarray}
U |\varphi|^2 -\mu -t = 0 \,\, \Longrightarrow \,\, \varphi_{GP} =
\sqrt { \frac{\mu+t}{U}}.
\end{eqnarray}
The quantity $|\varphi|^2$ is the density in the uniform system,
and is related to the chemical potential. Conversely, for a given
density $|\varphi|^2$, the chemical potential is given by

\begin{eqnarray}
\mu=U|\varphi|^2-t,  \label{condmu}
\end{eqnarray}
namely the kinetic energy plus interaction energy. Strictly speaking
the kinetic energy of the condensate is zero, and $t$ is just a shift
due to the choice of the zero of energy. This approach does not
include any phase transition, because the SF fraction $|\varphi|^2$ is
always equal to the total density.

To diagonalize the second order terms in the Hamiltonian ($H_2$),
we perform a transformation to momentum space

\begin{eqnarray}
{\tilde a}_i&=& 
\frac{1}{N} \sum_{\bf q} e^{-i{\bf q}\cdot {\bf r_i}}{\tilde a}_{\bf q}, \\
{\tilde a}_i^{\dag}&=& 
\frac{1}{N} \sum_{\bf q} e^{i{\bf q} \cdot {\bf r_i}}{\tilde a}_{\bf q}^{\dag},
\end{eqnarray}
so that, finally it reads

\begin{eqnarray}
&H_2& = - \frac{1}{2} \sum_{\bf q} \left[ 2U |\varphi|^2 -\mu +
\epsilon({\bf q}) \right] +
\\
&+& \frac{1}{2} \sum_{\bf q} \left[ \left(2U|\varphi|^2 - \mu +
\epsilon({\bf q}) \right) \left( {\tilde a}_{\bf q}^{\dag} {\tilde
a}_{\bf q}
+ {\tilde a}_{-{\bf q}}{\tilde a}_{-{\bf q}}^{\dag} \right) + \right. \nonumber \\
&& \left. + U \varphi^2 {\tilde a}_{\bf q}^{\dag} {\tilde
a}_{-{\bf q}}^{\dag} + U \varphi^{*2}{\tilde a}_{-{\bf q}} {\tilde
a}_{\bf q} \right],  \nonumber
\end{eqnarray}
where, as before, $\epsilon({\bf q})=-(2t/z)\sum_i\cos(q_i)$.  Then,
we apply the Bogoliubov transformation which diagonalizes $H_2$

\begin{eqnarray}
\left(
\begin{array}{c}
{\tilde a}_{\bf q} \\
{\tilde a}_{-{\bf q}}^{\dag} \\
\end{array}
\right)= \left(
\begin{array}{c}
u_{\bf q} b_{\bf q} + v^*_{-{\bf q}}b_{-{\bf q}}^{\dag} \\
u^*_{-{\bf q}} b_{-{\bf q}}^{\dag} + v_{\bf q} b_{\bf q}
\end{array}
\right),  \label{unif}
\end{eqnarray}
with the additional condition $|u_{\bf q}|^2 - |v_{-{\bf q}}|^2
=1$ to preserve the commutation relations. This is equivalent to
the Bogoliubov equations

\begin{eqnarray}
(\mathcal{L}_{\bf q} -\hbar \omega_{\bf q}) u_{\bf q}+ U \varphi^2 v_{\bf q} = 0,  \nonumber \\
(\mathcal{L}_{\bf q} + \hbar \omega_{\bf q}) v_{\bf q} + U
\varphi^{*2} u_{\bf q} = 0, \nonumber
\end{eqnarray}
where $\mathcal{L}_{\bf q}=U|\varphi|^2 + (4t/z) \sum_i
\sin^2(q_id/2)$.

The solution for the Bogoliubov spectrum and the Bogoliubov amplitudes
is given by

\begin{eqnarray}
\hbar \omega_{\bf q} = \sqrt{\frac{4t}{z} \sum_i
\sin^2\left(\frac{q_i}{2}\right) \left[\frac{4t}{z} \sum_i
\sin^2\left(\frac{q_i}{2}\right) + 2 U
|\varphi|^2 \right]}, \nonumber \\ \\
u_{\bf q}+v_{\bf q}=\sqrt{\frac{\mathcal{L}_{\bf q}-U\varphi^2}{\hbar \omega_{\bf q}}} =
 \sqrt{\frac{(4t/z) \sum_i
\sin^2(q_i/2)}{\hbar \omega_{\bf q}}}, \hspace*{0.5cm}\\
u_{\bf q}-v_{\bf q}=\sqrt{\frac{\hbar \omega_{\bf q}}
{\mathcal{L}_{\bf q}-U\varphi^2}} =\sqrt{\frac{ \hbar \omega_{\bf
q}}{(4t/z) \sum_i \sin^2(q_i/2)}}. \hspace*{0.5cm}
\end{eqnarray}
For ${\bf q}\to 0$, the spectrum shows a linear behavior in $|{\bf
q}|$, $\hbar \omega_{\bf q} \approx |{\bf q}| \sqrt{ (2t/z) U
|\varphi|^2}$ with sound velocity $c= \sqrt{ (2t/z) U
|\varphi|^2]}$.

In the Bogoliubov theory the momentum distribution is given by

\begin{eqnarray}
n({\bf q})= |\varphi|^2 \delta_{{\bf q},0} + |v_{-{\bf q}}|^2.
\end{eqnarray}
It is evident that starting from the fact that $|\varphi|^2$ equals
the total density in the uniform system, the integral of the momentum
distribution $\int n({\bf q}) d{\bf q}$ will exceed this value by the
depletion $n_D=\int |v({\bf q})|^2 d{\bf q}$. In the limit of validity
of the Bogoliubov approach, this quantity is very small.

\section{Analytic calculation of the momentum distribution in
the Mott regime in the RPA approximation}

\label{app3}

In the RPA approximation, the momentum distribution in the MI regime
can be calculated analytically. In this appendix, we will present the
results for 1, 2, and 3D systems, with the aim of pointing out that
close to the phase transition quantum fluctuation play a major role
and are not correctly taken into account by RPA.

In the MI regime, the RPA Green's function in Eq.~(\ref{an_sol}) can
be written as

\begin{eqnarray}
G_{MI}({\bf q},\omega)=\frac{1}{2\pi} \frac{A_{11}(\omega)}{1 -
\epsilon({\bf q}) A_{11}(\omega)} ,  \label{gmott}
\end{eqnarray}
where $A_{11}$ has been defined in Eq.~(\ref{a11_gen}),
$A_{12}=A_{21}=0$ in the MI, and  $\epsilon({\bf
q})=-(2t/z)\sum_i\cos(q_i)$ as before.

In particular for the MI at density $n$ (i.e. where the on-site
ground state is $|0\rangle=n$), we get

\begin{eqnarray}
A_{11}(\omega)= \frac{n+1}{\omega-(E_{n+1}-E_n)}-
\frac{n}{\omega+(E_{n-1}-E_n)}, \label{a11}
\end{eqnarray}
(with $E_n=-n\mu +(U/2) n(n-1)$) which obviously coincides a part from
a factor $1/2\pi$ with the Green's function for the MI at zero
tunneling previously introduced in Eq.~(\ref{gmott_t0}).

By using definitions in Eqs.~(\ref{gmott},\ref{a11}) above and
defining $\Delta_{\pm}=E_{n\pm1}-E_n$, it is straightforward to see
that the Green's function takes the form

\begin{eqnarray}
&&G_{MI}= \\
&&=\frac{1}{2\pi}  \frac{(n+1)(\omega+\Delta_-) - n
(\omega-\Delta_+)}
{(\omega-\Delta_+)(\omega+\Delta_-)-\epsilon({\bf q}) [n \omega +
(n+2)\Delta_-] } , \nonumber
\end{eqnarray}
whose poles $\omega_{\pm}$ can be calculated analytically and have a
momentum dependence due to the kinetic energy $\epsilon({\bf q})$ in
Eq.~(\ref{gmott}). Hence, the Green's function close to the poles
($\omega \approx \omega_\pm$), is approximately

\begin{eqnarray}
G_{MI}\approx \frac{1}{2\pi} \frac{(n+1) (\omega_\pm + \Delta_-)-n
(\omega_\pm -\Delta_+)}{ (\omega_\pm -
\omega_\mp)(\omega-\omega_\pm)}.
\end{eqnarray}
Consequently, the momentum distribution reads

\begin{eqnarray}
n({\bf q}) = - \frac{(n+1)(\omega_- + \Delta_-) - n (\omega_- -
\Delta_+)}{\omega_- - \omega_+}. \label{nqmottan}
\end{eqnarray}
In the deep MI ($t=0$), $\omega_\pm = \pm \Delta_\pm$ and the momentum
distribution is simply given by $n({\bf q})=n$, which integrated over
the allowed momenta (equal to the total number of wells) gives the
total number of atoms. At finite tunneling, the kinetic energy
$\epsilon({\bf q})$ gives a modulation to the momentum distribution.

\begin{center}
\begin{figure}[ht!]
\includegraphics[width=0.75\linewidth]{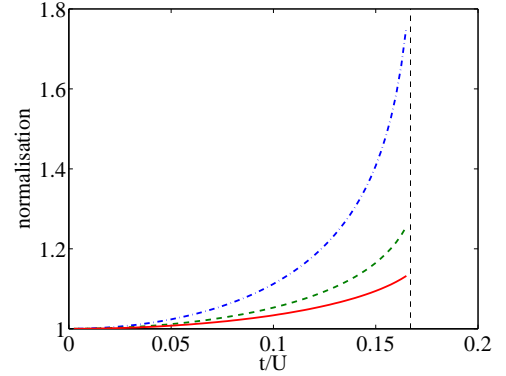}
\caption{Normalization of $n(q)$ for $\mu/U=0.5$ as a function of
$t/U$ in dimensions 1D (blue dashed-dotted line), 2D (green dashed
line) and 3D (red full line). The thin vertical line indicates the
phase transition.} \label{normalization}
\end{figure}
\end{center}

To find the normalization of $n({\bf q})$, we integrate numerically
the analytic expression in Eq.~(\ref{nqmottan}) for different
dimensions. The result is shown in Fig.~\ref{normalization} for the MI
with $n=1$, and clearly shows an increase of the normalization toward
the phase transition. We attribute this effect to the fact that
quantum fluctuations are not completely taken into account in this
approach. This interpretation is confirmed by the fact that this
effect diminishes by increasing the dimensionality of the system.

It is interesting to note that while the normalization of the momentum
distribution is violated, the sum-rule is perfectly satisfied. The
strengths at positive and negative energy are respectively given by

\begin{eqnarray}
S_{\pm}=  \frac{(n+1) (\omega_\pm + \Delta_-)- n (\omega_\pm -
\Delta_+)}{\omega_\pm - \omega_\mp},
\end{eqnarray}
such that $S_++S_-=1$.

\end{document}